\documentclass[english,submission,copyright,creativecommons]{eptcs}
\usepackage[T1]{fontenc}
\usepackage[latin9]{inputenc}
\usepackage{array}
\usepackage{multirow}
\usepackage{graphicx}
\usepackage{eurosym}
\usepackage{threeparttable}



\usepackage{babel}

\author{Chris Chapman
\institute{Royal Military College of Canada,\\
Electrical and Computer Engineering Department, Kingston, Canada}
\email{chrisajchapman@hotmail.com}
\and
Scott Knight
\institute{Royal Military College of Canada,\\
Electrical and Computer Engineering Department, Kingston, Canada}
\email{knight-s@rmc.ca}
\and
Tom Dean
\institute{Queen's University, Electrical and Computer Engineering
  Department,\\ Kingston, Canada}
\email{tom.dean@queensu.ca}
}

\def\titlerunning{USBcat -- Towards an Intrusion Surveillance Toolset}
\def\authorrunning{C. Chapman, S. Knight, T. Dean}

\begin{document}

\title{USBcat -- Towards an Intrusion Surveillance Toolset}
\maketitle
\begin{abstract}
This paper identifies an intrusion surveillance framework which
provides an analyst with the ability to investigate and monitor
cyber-attacks in a covert manner. Where cyber-attacks are perpetrated
for the purposes of espionage the ability to understand an adversary's
techniques and objectives are an important element in network and
computer security. With the appropriate toolset, security
investigators would be permitted to perform both live and stealthy
counter-intelligence operations by observing the behaviour and
communications of the intruder. Subsequently a more complete picture
of the attacker's identity, objectives, capabilities, and infiltration
could be formulated than is possible with present technologies.\\
This research focused on developing an extensible framework to permit
the covert investigation of malware. Additionally, a Universal Serial
Bus (USB) Mass Storage Device (MSD) based covert channel was designed
to enable remote command and control of the framework. The work was
validated through the design, implementation and testing of a toolset.
\end{abstract}

\section{Introduction}
\label{sec:Introduction}

Computer network defence is enhanced where there is an understanding
and appreciation of an adversary's strategies and objectives. This is
especially important when the network has been targeted by a
sophisticated, persistent attacker. A challenge faced by network
defenders is the ability to conduct an investigation of a compromised
machine without alerting the cyber-intruder of the defender's
activities. While tools such as honeypots exist to permit in-depth
analysis of malware they do not necessarily provide for live analysis
and interaction with the malware. The intent of our framework is to
provide an extensible platform which enables covert and dynamic
investigation of malware in real-time, in order to identify the
attacker. A stealthy investigative capability is of particular
importance when attempting to gain intelligence on well funded and
highly skilled adversaries such as foreign intelligence agencies and
criminal elements seeking unauthorized access to private and
classified networks. Once an intruder has breached the perimeter
security, subsequent actions they may take include further
infiltration into the network as well as the establishment of covert
channels to exfiltrate information from the organisation. Information
Technology (IT) security responses commonly seek to quickly isolate
the threat from the network and perform an off-line investigation.
While adequate for generic malware, this method of investigation will
likely alert a more advanced attacker, permitting them the opportunity
to conceal both the origin and depth of the attack. Equipped with an
appropriate toolset, security investigators would then be able to
perform live counter-intelligence operations by observing the
behaviour, actions, and communications of the intruder. These
counter-intelligence operations then provide a more complete picture
of the attacker's intentions, capabilities, depth of infiltration, as
well as their potential identity \cite{1}.

\subsection{Motivating Scenario}

To identify the capabilities required by an analyst in an
investigation we considered a scenario in which an analyst is
investigating a compromised machine where espionage activities are
suspected. In order to conduct such an investigation the analyst will
require the ability to watch the intruder in action. Any use of local
Input/Output (IO) devices, such as keyboard, mouse, and monitor are
considered off limits in our solution given that their use would be
highly visible to the intruder. This implies that the analyst needs
the ability to stealthily effect and observe actions taking place on
the compromised host from a remote machine. These constraints were
used to guide the course of our research.

\subsection{Aim}

This paper will proceed as follows. Section \ref{sec:2} provides
background research that guided the development of our framework.
Section \ref{sec:3} discusses the design of our framework and covert
channel. Section \ref{sec:4} presents the validation of our toolset.
Section \ref{sec:5} presents concluding remarks. Future work is
addressed in Section \ref{sec:6}.

\section{Background Research}
\label{sec:2}

In the introduction we presented the motivation for the development of
a toolset capable of discretely monitoring the actions of an intruder.
In order to be stealthy our toolset must be able to execute on the
compromised machine in a manner which is difficult for even a
sophisticated attacker to detect. In order to advance towards these
objectives we will require deeper insight into three areas of
research, these include:

\begin{itemize}
\item Hiding the toolset on the compromised machine,
\item Identification of the tools and capabilities the analyst will require, and
\item Interacting with the toolset in a covert manner.
\end{itemize}

To conceal the presence of our toolset on the compromised machine,
rootkit technologies were explored in order to identify techniques and
capabilities that would offer our toolset stealth on the compromised
host. Additionally, an operational scenario was explored for the
purposes of identifying tools and capabilities pertinent to carrying
out an investigation of a compromised machine. Also, the analyst must
be able to interact with the toolset in order to carry out his
investigation. Our research identified that a Mass Storage Device
(MSD) based covert channel would serve our purposes well. We will
therefore elaborate on MSD protocol attributes which supported this
design decision.

\subsection{Hiding The Toolset}

Rootkits provide the capability of concealing the truth about what is
actually taking place within the operating system and were therefore
considered for the purposes of hiding our toolset on the compromised
machine. The intent was to identify a rootkit which adequately
addressed the three criteria: stealth, the semantic gap, and data
exfiltration. Ultimately a novel Kernel Mode (KM) based rootkit known
as Dark Knight \cite{2} was identified as good platform through which
to launch and conceal the presence of our toolset. Dark Knight is a
developmental rootkit which has been designed for the purposes of IT
security research. This rootkit leverages Asynchronous Procedure Calls
(APC) within the Microsoft Windows Operating System (OS) as a means to
provide stealth and conceal the presence of executable code \cite{2}.

Of particular interest is the ability for APCs to inject code into
processes across the entire OS, providing access to variables and
objects within the virtual memory space of the targeted process. The
ability to target malware concealing its presence in a legitimate
process is therefore enabled.

As will be discussed later, our framework does entail the launching of
processes and threads. Dark Knight does not in itself conceal the
presence of these artifacts and while other traditional KM rootkit
techniques may be invoked for this purpose, such activity is
considered outside the scope of our research.

\subsection{Tools and Capabilities Identification}

A key component of our framework must be its ability to facilitate an
analyst's investigation of a compromised host. Of particular interest
is investigation of the processes, executables, files and network
connections associated with the malicious activity. Tools critical to
this investigation included those that are native to the Windows
environment. These include the capability to navigate the file system.
Additionally tools such as netstat to list network connections and
tasklist to list active processes are also critical.

There also exists many analytical tools that are not natively found on
a Windows machine but are very useful in investigating process
activity. These include memory dumping tools such as Userdump \cite{3}
and DumpIt \cite{4}. Given that many analytical tools already exist
for the purpose of analyzing software it will be a goal of our
framework to incorporate and leverage existing capabilities. This will
in turn ensure our framework is applicable across a wide variety of
malware detection and analysis scenarios.

\subsection{Command and Control}

For the purposes of enabling covert communications within our
framework a novel use of an USB MSD channel was developed and
implemented. Previous research has demonstrated the use of USB based
devices for hosting covert channels \cite{5}. A covert channel
implemented over an MSD was selected for its potential to enable
high-bandwidth transmissions and support our requirement to off-load
log/data files from the compromised machine. Intuitively, a covert MSD
channel also offers enhanced stealth over a network based covert
channel, in that communications cannot be observed by a third party
residing on the network.

\begin{figure*}[hbtp]
\begin{centering}
\includegraphics[scale=0.6]{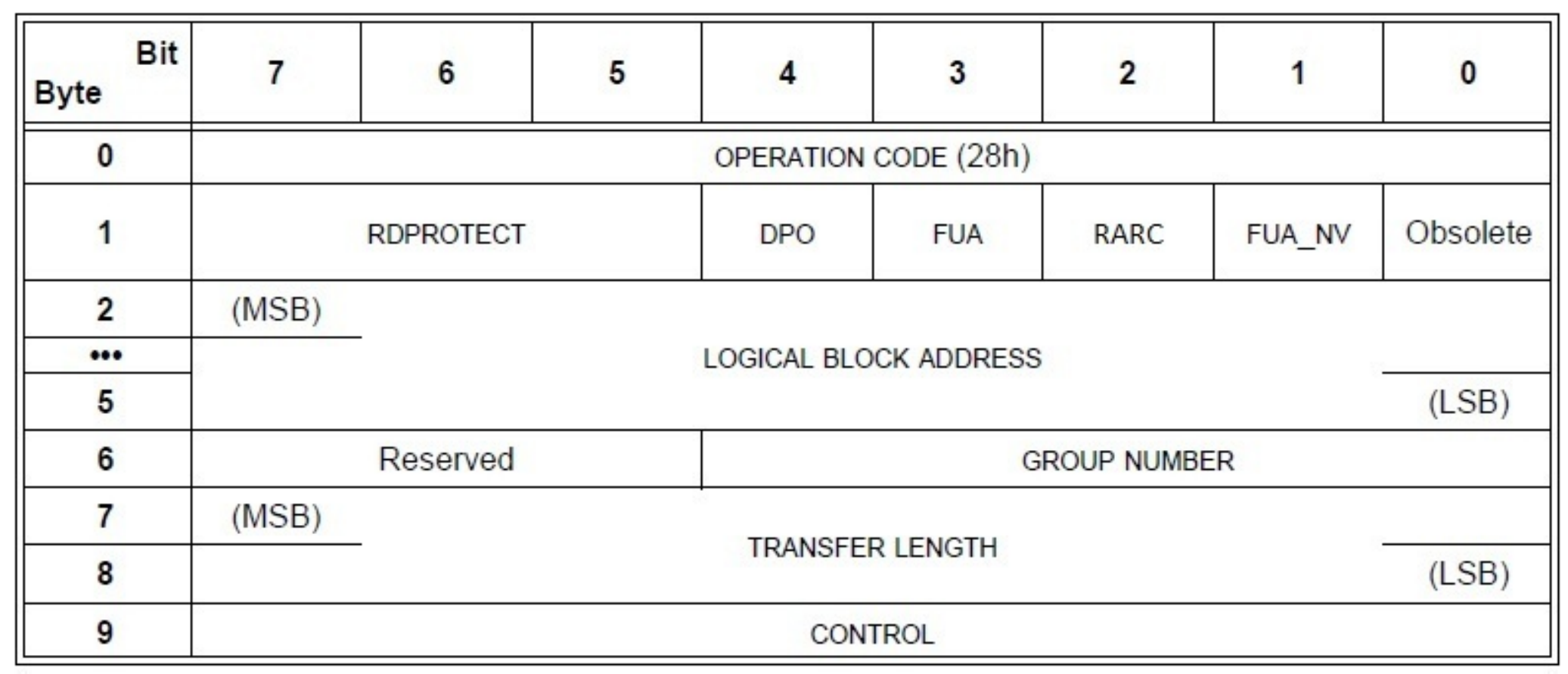}
\par\end{centering}
\caption{READ Frame (from \cite{6})\label{fig:1}}
\end{figure*}

A host makes use of the Small Computer Systems Interface (SCSI)
protocol to communicate and initiate data transfers with an MSD. Of
particular interest to us in our research with respect to developing a
covert command and control channel are the abilities to communicate
discretely and carry out the transfer of large volumes of data. For
example, the case where log files are off loaded from a compromised
machine. For these reasons our focus was directed to the following
SCSI commands: READ, WRITE and TEST UNIT READY. The READ and WRITE as
their names suggest are the commands for reading from and storing
information to the MSD. Data on an MSD is referenced using a logical
block address (LBA) scheme. Read and write requests therefore include
in their request the starting LBA they intend to access in addition to
the number of blocks that are being requested and/or written. Figure
\ref{fig:1} depicts the fields in a READ command block. The first
field carries the operation code of the command which is 28h in the
case of a READ request. The address of the LBA occupies bytes 2
through 5 and the transfer length (number of LBAs requested) occupies
bytes 7 and 8. A WRITE command block is similar in nature with the
exception that its operation is designated by 2Ah.

\begin{figure*}[hbtp]
\begin{centering}
\includegraphics[scale=0.6]{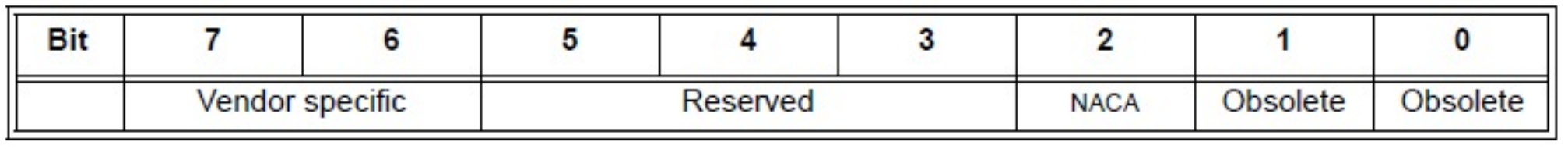}
\par\end{centering}
\caption{Control Byte (from \cite{7})\label{fig:2}}
\end{figure*}

In the interest of developing a covert channel the Control field (byte
9) was inspected further in order to identify any settings which might
be open to subversion. As identified in Figure \ref{fig:2}, bit 6 and
7 are to contain vendor specific information. In our development
environment it was identified that these bits are not in use and we
are able to set them as we saw fit without adversely affecting
communications. With these bits the potential exists to distinguish
between frames associated with a covert channel and normal host/MSD
communications.

Further analysis of typical host/MSD communications revealed the
presence of a heart beat type signal used to determine if the device
is ready for use. This signal is the TEST UNITY READY signal and is
sent at a frequency of approximately twice per second. Given that our
toolset will need to be in regular communication with the analyst for
the purposes of receiving command and control signals, the TEST UNIT
READY command provides a good candidate signal for subversion. By
leveraging it we would be able to hide within the noise of normal
host/device communications.

\section{Framework Design}
\label{sec:3}

\begin{figure*}[hbtp]
\begin{centering}
\includegraphics[scale=0.9]{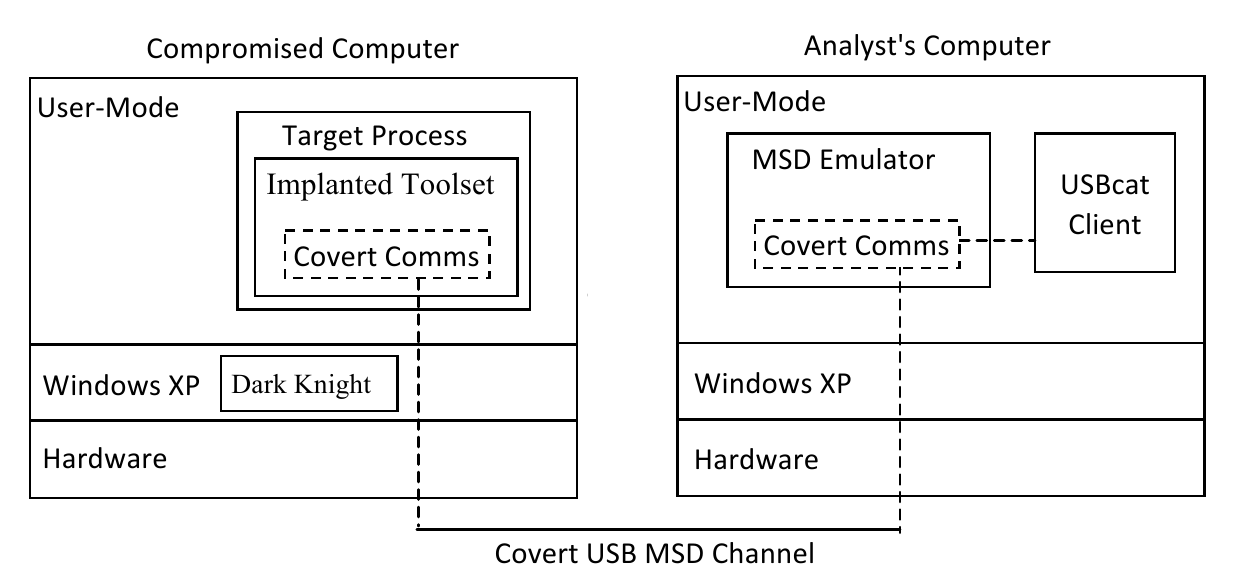}
\par\end{centering}
\caption{Surveillance Framework High-Level Design\label{fig:3}}
\end{figure*}

Our design consists of components on two separate computers, the
compromised computer and the analyst's computer from which commands
are issued in order to carry out the investigation. Five major
components form the basis of our architecture as depicted in Figure
\ref{fig:3}, these include an implanted toolset, a covert
communication capability, an MSD emulator, the Dark Knight rootkit and
an USBcat client. This design models a client-server architecture,
with the implanted toolset acting as a server and permitting the
USBcat client (analyst) to establish investigative sessions on the
compromised machine.

The choice of the name USBcat is based on our use of an USB channel
for covert communications in addition to our integration of a netcat
\cite{8} like framework to execute analytical tools on the compromised
machine. Detail regarding our netcat style framework will be presented
within this section.

The USBcat client will act as the analyst's main interface for
carrying out command and control of the framework. The MSD emulator
will emulate the existence of an MSD (eg. a thumb drive) which will be
used to develop a covert communications channel. The Dark Knight
rootkit will serve to conceal the presence of the implanted toolset on
the compromised machine, in addition to providing a user-space
foothold on the compromised machine. The implanted toolset itself will
manage sessions in the target and provide a scalable architecture to
allow multiple concurrent client sessions to be actively running
investigative programs and tools. The covert communications capability
will consist of a communications module leveraged by the implanted
toolset on the compromised computer in addition to functionality
incorporated into the MSD emulator on the analyst's computer.

\subsection{Implanted Toolset}

\begin{figure*}[hbtp]
\begin{centering}
\includegraphics[scale=0.9]{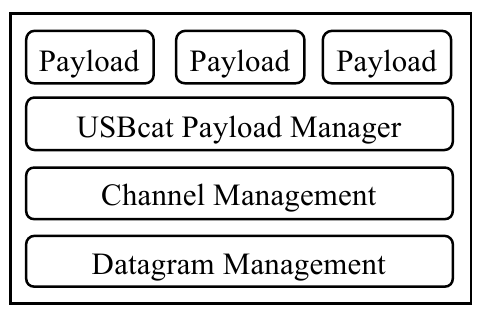}
\par\end{centering}
\caption{Surveillance Framework High-Level Design\label{fig:4}}
\end{figure*}

Our design seeks to produce an architecture which is modular,
scalable, and architecturally independent from the communication
channel on which it resides. These attributes are visible in the
modular nature of our implanted toolset (Figure \ref{fig:4}). The
implanted toolset is the portion of our design that resides entirely
on the compromised machine. This toolset was implemented as dynamic
link libraries (DLLs) that are injected by the Dark Knight rootkit
into a target process on the compromised machine.

The implanted toolset can be subdivided into a datagram management, a
channel management and a payload management component, as depicted in
Figure \ref{fig:4}. The upper layer module is entitled the USBcat
payload manager and is responsible for launching payloads as requested
by the analyst. Payloads are analysis tools that the analyst can
launch on the compromised machine.\\

\noindent {\bf USBcat Payload Manager}. Our implanted toolset is
capable of executing diagnostic commands on behalf of the analyst.
This includes both native operating system commands and utilities in
addition to custom system analysis tools. Rather than reimplement
system diagnostic tools that already exist we chose to implement a
mechanism that would make use of existing utilities. Our solution to
meet this requirement was to design an architecture which models that
of the netcat \cite{8} application. Of interest to us is the technique
used by the netcat architecture to spawn a child process and through
the use of pipe streams direct commands to the child process as well
receive the results of those commands. In this way a netcat listener
session is able to relay commands received from a netcat client to the
spawned process (i.e. the diagnostic tool) and respond with the
results of those commands. By applying a similar design we are able to
launch and control analytical tools on the compromised machine. Where
analytical tools do not natively reside on the compromised machine our
toolset incorporates the ability to transfer the executable from the
analyst's machine for local execution.

\begin{figure*}[hbtp]
\begin{centering}
\includegraphics[scale=0.9]{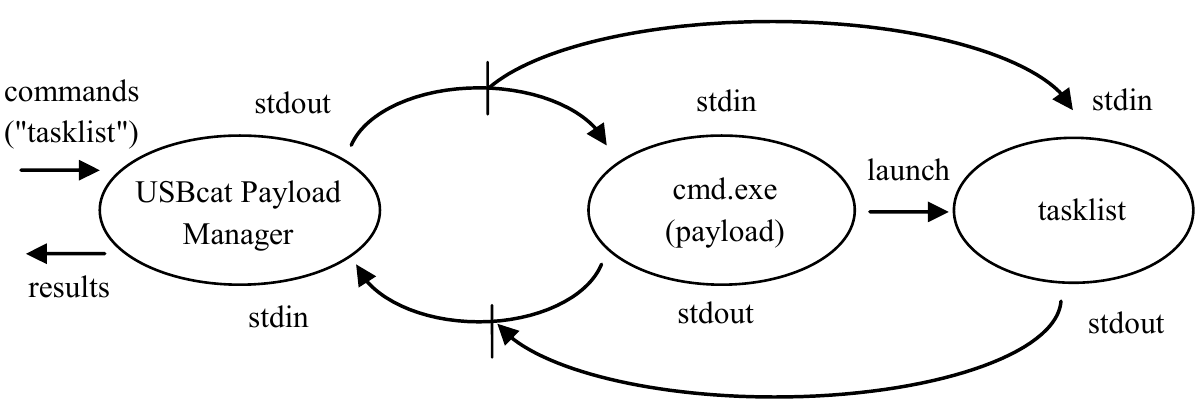}
\par\end{centering}
\caption{USBcat Payload Launching Behaviour\label{fig:5}}
\end{figure*}

The USBcat payload manager is used to initially spawn the payload
requested by the analyst. For the purposes of this explanation we will
consider the launching of the command shell process as depicted in
Figure \ref{fig:5}. Communications with this child process are
implemented with two threads of execution within the payload manager.
One thread waits for commands to be received from the channel and
subsequently issues a WriteFile() \cite{9} command which directs the
input into a shared pipe between the USBcat payload manager and the
command shell's stdin. While the other thread is responsible for
issuing ReadFile() \cite{10} commands in order to check the shared
pipe connected to the stdout of the command shell. The command shell
will then invoke the execution of commands and utilities it receives.
These commands and utilities are then themselves executed as child
processes of the command shell, thereby inheriting the input and
output channels of their parent. Output from this command/utility is
then directed back to the payload manager.\\

\noindent {\bf Channel and Datagram Management}. The architecture of
the USBcat implanted toolset is further supported by the channel and
datagram management modules. These modules offer a layer of
abstraction and separation from the actual implementation of the
covert channel. The channel management module provides scalability to
our solution by permitting multiple investigative sessions to coexist
and communicate with the analyst. The datagram management module is
responsible for implementing the underlying covert communication
channel between the implanted toolset and the USBcat client on the
analyst machine. This module implements communication through direct
reads and writes to blocks on the MSD.

\subsection{Covert Communications Capability}

Design elements for the covert communication between the implanted
toolset and the analyst will now be presented.\\

\noindent {\bf MSD Emulator}. A key component of our covert channel is
the ability to intercept and generate MSD communications at the
analyst's end of the channel. To facilitate the development of this
channel we made use of an USB device development environment
\cite{11}. Our module is designed to inspect all communications
received by the emulated MSD. Regular host to MSD communications will
make use of the standard MSD implementation code, whereas
communications identified as part of our covert channel will be
diverted to our built-in toolset.\\

\noindent {\bf Covert Communications: Polling and Data Transmission.}
In order to implement our covert channel the implanted toolset
requires the ability to be notified when the remote end of the channel
has a datagram ready to send. This ability was implemented as a
polling activity to be carried out by the implanted toolset. During
the course of our research we identified that the native behaviour of
a host machine with an attached MSD is to send it Test Unit Ready
signals at regular intervals (approximately twice per second). We
therefore implemented our own Test Unit Ready signals for the purposes
of polling the USBcat client for commands. The decision was made to
leverage the Test Unit Ready signal given the fact that we are seeking
to hide our channel within the existing noise of standard host/MSD
communications.

\begin{figure*}[hbtp]
\begin{centering}
\includegraphics[scale=1]{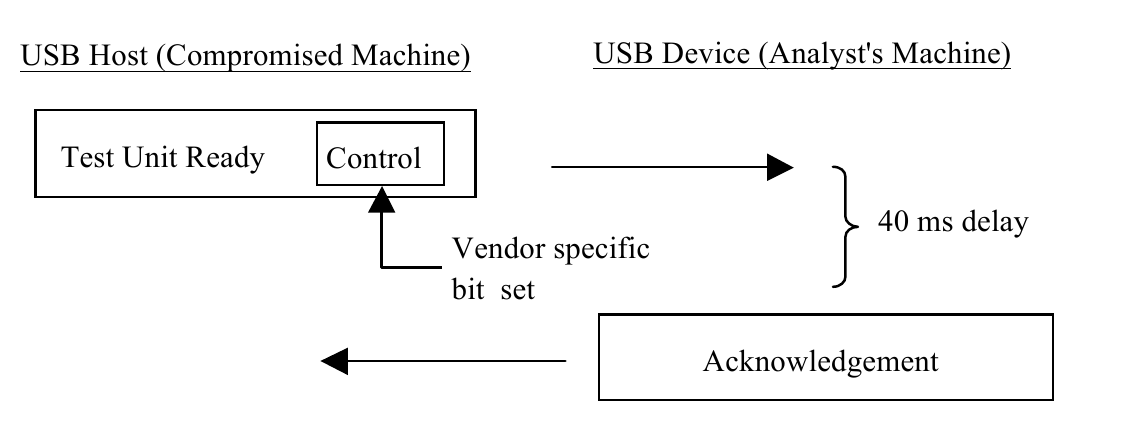}
\par\end{centering}
\caption{Covert Signal -- Polling Mechanism\label{fig:6}}
\end{figure*}

Given that there now exists multiple Test Unit Ready commands being
received by the MSD, our covert communication module in the MSD needs
a means to distinguish between regular host signals and our covert
polling signals. This is achieved by making use of the control byte in
the Test Unit Ready packet and setting one of the vendor specific
bits. The toolset at the analyst end of the channel is therefore able
to identify Test Unit Ready signals sent by the implanted toolset. The
covert communication module in the MSD is able to alert the implanted
toolset when the analyst has a command ready to send by delaying the
Test Unit Ready packet acknowledgement by a fixed amount of time as
depicted in Figure \ref{fig:6}. A delay of 40 msec was sufficient for
our purposes. This delay is identified by the implanted toolset, which
then issues a Read request to the analyst side toolset in order to
request the next available message.

To distinguish our covert channel Read requests from that of normal
MSD behaviour we again set a vendor specific bit. This custom Read
request then informs our analyst side toolset that the implanted
toolset is requesting a command from the analyst. The analyst end of
the channel then responds with the next available datagram. As results
of the analyst's commands become available from the implanted toolset
they are transmitted to the analyst with the use of Write requests. In
a similar manner a vendor specific bit is set in order to identify our
communications as part of the covert channel.

\section{Validation Activities}
\label{sec:4}

Validation of the USBcat toolset was carried out to determine if its
operation would be detectable by an adversary. Based on our results,
we argue that our toolset permits an investigation of malicious code
on a compromised to be carried out remotely and with a high degree of
stealth. Our validation also considered the extensible nature of our
toolset and its versatility for enabling investigations against the
broader context of malware compromises.

\begin{figure*}[hbtp]
\begin{centering}
\includegraphics[scale=1]{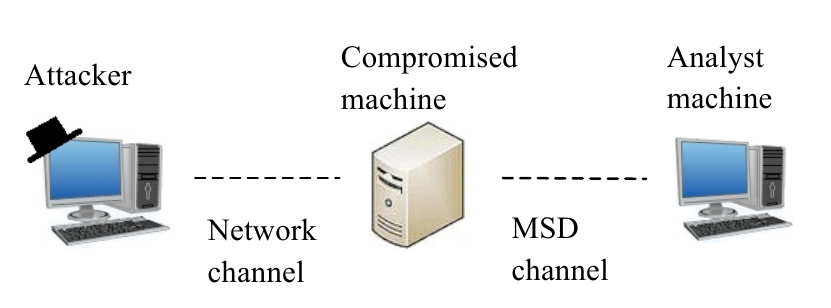}
\par\end{centering}
\caption{Validation Environment\label{fig:7}}
\end{figure*}

Our performance testing considered the performance impact on the
compromised machine from the perspective of both internal and external
time sources. That is, would an attacker notice or could he observe
changes in performance that would alert him to the defender's
activities. From the internal time source perspective we considered
the scenario where the adversary is attempting to detect the presence
of our toolset from a vantage point within the compromised machine.
Alternatively, for the external time source scenario we evaluate the
potential for the adversary to detect the performance impact of our
toolset from their attacking machine. Figure \ref{fig:7} depicts the
validation environment.

\subsection{Performance Measurements - Internal Time Source}

To conduct the internal time source measurements we made use of
Microsoft performance \cite{12} counters which monitor running
applications and system resources. Our motivation for the selection of
such counters was to assess the impact of our toolset on system
performance. These counters capture I/O activity in terms of file
system reads and writes carried out by our toolset, in addition to
measuring processor and memory usage. As our implanted toolset is
designed to execute within the context of an existing process on the
compromised machine, we must first select a target process. The target
process for our validation was the explorer.exe process. Explorer.exe
was selected given that it is both a native and privileged process
which executes within the Microsoft OS and is therefore also the
likely target of an adversary (for the injection of malicious code). A
total of six measurement scenarios were performed:

\begin{itemize}
\item Baseline 1 (idle)
\item Baseline 2 (end-user activity)
\item Toolset scenario 1 (USBcat injected but idle)
\item Toolset scenario 2 (processing "dir C:$\backslash$WINDOWS" at 3 second intervals)
\item Toolset scenario 3 (executing file transfers of 34 MB file)
\item Toolset scenario 4 (executing Userdump.exe)
\end{itemize}

The baseline scenarios provide us with a normal picture of the
compromised machine's performance without our toolset installed. The
first baseline represents the scenario where the compromised machine
is otherwise idle, while the second baseline incorporates end-user
activity with events such as document editing and web browsing
traffic. Our intent is to better understand how typical end-user
activity may further conceal the presence of our toolset.

In addition to the two baseline scenarios, four toolset scenarios were
evaluated. These scenarios represent typical activities that would be
expected to take place over the course of an investigation of a
compromised machine. Additionally, these scenarios represent an
increasing load on system resources of the compromised machine and
therefore provide insight into the impact of toolset activity and
intensity on system resources.

Upon comparing baseline 1 with toolset scenario 1 (Table \ref{tab:1})
there exists a number of metrics which fall outside of two standard
deviations and are therefore potentially suspicious to an attacker
concerned with identifying anomalous behaviour on the compromised
machine. However, as indicated in baseline 2 as end-user activity is
added to the machine, toolset activities cease to be anomalous and
very quickly become hidden in the noise of the system.

\begin{table*}[hbtp]
\begin{centering}
\includegraphics[scale=1]{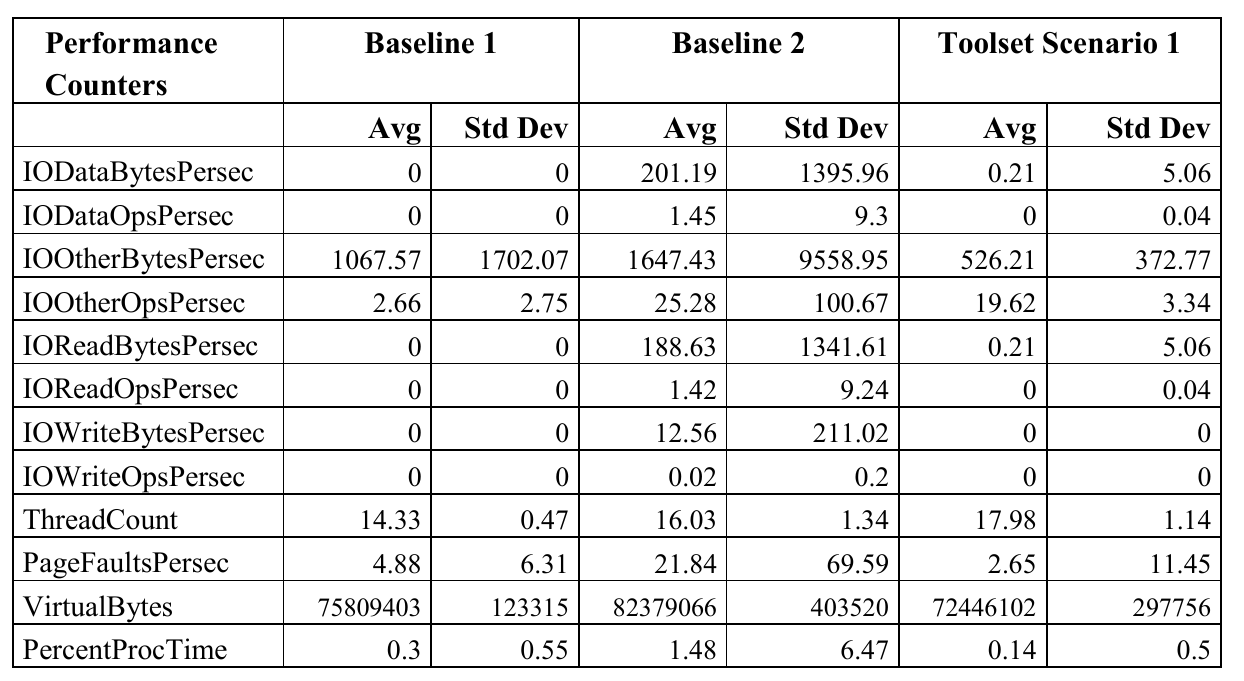}
\par\end{centering}
\caption{Internal Time Source Metrics -- Baselines \& Toolset Scenario 1\label{tab:1}}
\end{table*}

Toolset scenarios 2 and 3 (Table \ref{tab:2}) represent an increasing use of
recourses on the compromised machine. Where these activities begin to
stand out as anomalous the analyst's actions should be scaled down
such that metrics fall more in line with the rates identified in
baseline 2 (simulated end-user activity).

\begin{table*}[hbtp]
\begin{centering}
\includegraphics[scale=1]{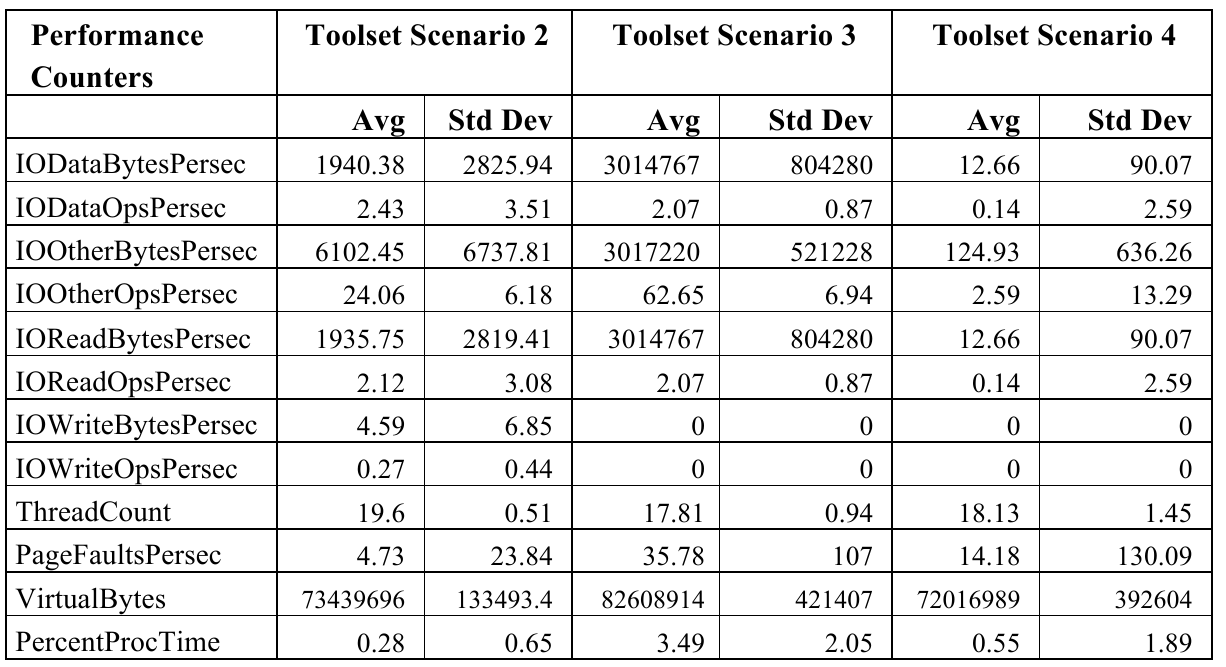}
\par\end{centering}
\caption{Internal Time Source -- Toolset Scenarios 2, 3 and 4\label{tab:2}}
\end{table*}

Toolset scenario 4 considered the impact of executing Microsoft's
Userdump executable as a payload in the compromised machine. During
the course of this scenario Userdump was observed temporarily locking
access to the process it is inspecting and therefore suspending all
other interaction with that process. Therefore diagnostic tools which
may significantly impact their target process must be used with care
in order to not disclose our toolset's presence.

The internal performance metrics gathered during this portion of the
validation support our argument that our toolset is sufficiently
stealthy and covert.

\subsection{Performance Measurements - External Time Source}

Performance metrics were also captured from an external time source.
This scenario is less intrusive and requires much less complexity to
execute from an attackers perspective as it can be conducted remotely.
This test involves repeated issuing of identical commands from the
attacker's workstation to the compromised machine. This then permits
measurements of the response time of each such command in order to
determine if the execution of our toolset is potentially visible to an
outside attacker.

To carry out this test we made use of the network depicted in Figure
\ref{fig:7}. Making use of tcpdump \cite{13} (on the attacker's
machine) we measured the round-trip time for the "dir
C:$\backslash$WINDOWS" command issued on the attacker's machine and
for the results of the command to be received. Test scenarios
incorporated the same toolset scenarios as those of the internal
performance measurements although during these tests we did not use a
baseline with end-user activity. Test results with no end-user
activity proved to be sufficiently stealthy in concealing the presence
of our toolset. For all scenarios "dir C:$\backslash$WINDOWS" was
issued 100 times from the attacker's computer at 3 second intervals.
Although, in toolset scenario 4 where Userdump was executing
concurrently our sample size was much smaller at just 14 samples.
Again, this was due to the fact that Userdump locks its target process
and therefore our target process explorer.exe was unable to respond in
a timely manner to the attacker's requests. The results of these tests
are listed in Table \ref{tab:3}.

\begin{table*}[hbtp]
\begin{centering}
\includegraphics[scale=1]{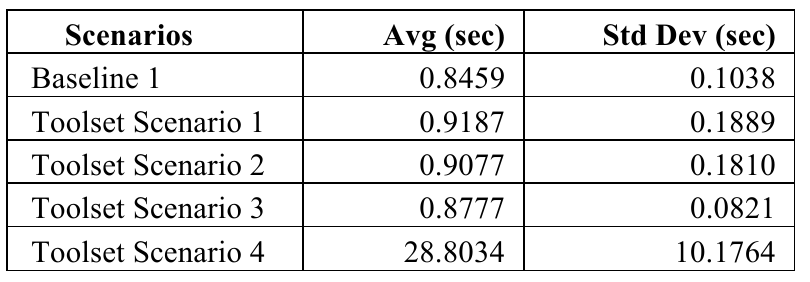}
\par\end{centering}
\caption{External Time Source Metrics\label{tab:3}}
\end{table*}

Toolset scenarios 1, 2 and 3 indicate only minor deviations from the
baseline in which no toolset is running. As these scenarios are well
within a single standard deviation of the baseline we are confident
that this method of instrumentation does not reveal the presence of
our toolset. By relying on measurements that are effectively measuring
round-trip times of individual commands the attacker will also be
impacted by fluctuations in response times resulting from other
traffic on the network. The combination of this network noise factor
in conjunction with the already minute timing variances depicted in
Table \ref{tab:3} would make it very difficult for external
performance testing to identify the presence of our toolset.

Alternatively, we see that in Scenario 4 response times to the
attacker's commands have increased greatly and are reminded that care
must be taken to ensure intrusive diagnostic tools are used
discretely.

\subsection{Validation Against the Broader Research Context}

Once an attacker gains a foothold on a compromised machine,
investigation of his actions will require standard tools and
capabilities such as those indicated in our design. By spawning a
command shell on the compromised machine and redirecting input and
output to the remote end of the toolset the analyst is able to launch
commands and carry out his investigation as if he were seated at the
compromised machine. The analyst can then use tools such as tasklist
and netstat to identify anomalous processes and network connections.

Following this initial assessment of a compromised machine often finer
grained and specialized toolsets will be needed to conduct a more
thorough investigation. Our toolset provides the capability to
transfer analysis tools to the compromised machine in a covert manner.
Of particular significance is that the use of console based analysis
tools require no modifications in order to be supported within our
toolset framework.

\section{Future Work}
\label{sec:5}

Current research opportunities will now be proposed. There are several
ways in which this research can be extended.

\subsection{Payload to Capture Stdin and Stdout}

The ability to see commands issued by the attacker to malware is a
very valuable capability towards identifying the attacker's overall
motivations and intent. Hooking stdin and stdout has commonly been
accomplished through filtering ReadFile and WriteFile system calls
within kernel space \cite{14}. While effective, a potentially more
novel technique might make use of APCs to filter system calls, and the
incorporation of this capability as a payload within our USBcat
framework. Given that APCs run at a higher Interrupt Request Level
(IRQL) than normal code \cite{15} they could be used to both detect
and filter ReadFile and WriteFile system calls in order to intercept
malicious command and control signals.

\subsection{Improved Covert Channel}

Our covert channel relies on polling behaviour in order to implement
our command and control channel. The implanted toolset polls for
commands by leveraging the Test Unit Ready signal. Rather than
carrying out polling activity a potentially more novel solution would
be to effectively leave the channel open allowing the analyst's end of
the channel to respond with command signals once they are ready
\cite{16}. Although, the serial nature of MSD communications may prove
to be a limiting factor to such a technique.

\subsection{Alternate Implanted Toolset Format}

Alexander identifies several types of code formats which the Dark
Knight rootkit is capable of injecting into user-mode threads
\cite{2}. When injecting code into a foreign process, code remapping
is required to correctly insert the code into the virtual memory
address space of this process. DLLs natively incorporate code
remapping capabilities in their executable, whereas the injection of
shellcode into a process requires the manual calculation of offsets.
The trade off is that DLLs provide a larger distinguishable footprint
in memory. Therefore implementation of the implanted toolset with the
use of shellcode would enhance its stealth.

\section{Conclusion}
\label{sec:6}

In our introduction we identified the strategic importance of
monitoring cyber espionage being conducted against government and
corporate infrastructure. During such attacks it is often of greater
strategic advantage to observe an intruder in order to gain enhanced
insight into their identity, objectives, strategy and infiltration
than to break contact and remove the attacker, effectively alerting
them of their detection. We therefore developed a framework to provide
such a stealthy investigative capability. The extensible nature of our
framework permits a wide variety of diagnostic tools to be seamlessly
executed on the compromised machine. Additionally, to maintain the
stealth of our investigation a novel covert USB MSD channel was
developed to permit remote stealthy command and control of our
toolset.

\bibliographystyle{eptcs}

\end{document}